\documentclass[doublespacing]{elsart}
\usepackage{graphicx}
\usepackage{amssymb}

\begin{document}

\begin{frontmatter}

\title{Spin Effects in a Quantum Ring}
\author[1]{T. Ihn}
\ead{ihn@phys.ethz.ch}
\author[1]{A. Fuhrer}
\author[1]{K. Ensslin}
\author[2]{W. Wegscheider}
\author[3]{M. Bichler}
\address[1]{Solid State Physics Laboratory, ETH Zurich, 8093 Zurich, Switzerland}
\address[2]{Institut f\"ur Experimentelle und Angewandte Physik, Universit\"at Regensburg, 93040 Regensburg, Germany}
\address[3]{Walter Schottky Institut, Technische Universit\"at M\"unchen, Am Coulombwall, 85748 Garching, Germany}
\begin{abstract}
Recent experiments are reviewed that explore the spin states of a ring-shaped many-electron quantum dot. Coulomb-blockade spectroscopy is used to access the spin degree of freedom. The Zeeman effect observed for states with successive electron number allows to select possible sequences of spin ground states of the ring. Spin-paired orbital levels can be identified by probing their response to magnetic fields normal to the plane of the ring and electric fields caused by suitable gate voltages. This narrows down the choice of ground-state spin sequences. A gate-controlled singlet--triplet transition is identified and the size of the exchange interaction matrix element is determined. 
\end{abstract}

\end{frontmatter}

In recent years, the vision of using quantum dots in the Coulomb blockade regime as realizations of quantum bits (qubits) within quantum information processing schemes has become a strong motivation for research on these systems.
Spin states in dots lend themselves for
the realization of spin qubits. It has been demonstrated theoretically \cite{loss} that experimental control over a time-dependent spin Hamiltonian of the form
\begin{equation}
\hat{H} = \sum_{i,j(<i)}J_{ij}(t)\vec{S}_i\vec{S}_j + \sum_i\mu_\mathrm{B}g_i(t)\vec{B}_i(t)\vec{S}_i
\label{ham}
\end{equation}
would allow the realization of all quantum logic operations required for quantum information processing. In eq.~(\ref{ham}), the first term describes the exchange interaction between pairs of spins $(\vec{S}_i,\vec{S}_j)$ with the exchange coupling constants $J_{ij}(t)$ and the second term describes the Zeeman splitting of spin states $\vec{S}_i$ in an external magnetic field $\vec{B}_i(t)$ weighted with the Land\'{e}-factor $g_i(t)$. Apart from its importance for qubit applications, the Hamiltonian (\ref{ham}) describes fundamental spin physics present in any quantum dot with negligible spin-orbit interaction. Owing to the fact that the spin states of quantum dots have been experimentally far less investigated than orbital (charge) states,
a large number of experiments were reported in recent years \cite{duncan00,folk01,lindemann02,hanson03,rokhinson01,ciorga02,luscher01,ihn03,tarucha00,ciorga01,kyriakidis02,fuhrer03,ono02,huttel03,pioroladriere03,ciorga03}.
Among them are studies of the Zeeman splitting in a parallel magnetic field \cite{duncan00,folk01,lindemann02,hanson03}, ground state spins in a perpendicular field \cite{rokhinson01,ciorga02}, spin pairing \cite{luscher01,ihn03}, singlet--triplet transitions as a function of magnetic field \cite{tarucha00,ciorga01,kyriakidis02,ciorga03} and gate voltages \cite{fuhrer03}, and spin-blockade effects \cite{ciorga02,ono02,huttel03}.
Many of these studies focus on single or coupled few-electron quantum dots \cite{hanson03,pioroladriere03,ciorga03}, in which the physics described by the above Hamiltonian is expected to appear in its simplest form.

In this paper, we focus on a Coulomb-blockaded many-electron quantum ring. Although many-electron systems are typically analyzed with statistical approaches \cite{alhassid00}, we were able to show previously that the level spectrum of high quality many-electron quantum ring structures in the Coulomb blockade regime can be understood in great detail \cite{fuhrer01}. Here we report experiments that exhibit the physics of the Hamiltonian (\ref{ham}), namely, the Zeeman effect and exchange interaction effects in this system. Furthermore, we demonstrate how a singlet--triplet transition based on exchange can be induced using appropriate gate voltages \cite{fuhrer03}.

The ring-shaped quantum dot sample has been fabricated on an AlGaAs-GaAs heterostructure containing a two-dimensional electron gas (2DEG) 34~nm below the surface, with density  $5\times 10^{15}$~m$^{-2}$ and  mobility  90~m$^2$/Vs at 4.2 Kelvin. The quantum ring has been defined by room temperature local anodic oxidation with a scanning force microscope (SFM) which allows to write oxide lines on a semiconductor surface that locally deplete the 2DEG underneath \cite{fuhrer02}. Figure~\ref{fig2}a shows an SFM image of the quantum ring with average radius $r_0=132$~{nm} containing of the order of 100 electrons. The number of electrons on the ring is tuned by applying voltages $V_\mathrm{pg1}$ and $V_\mathrm{pg2}$ to the plunger gates. Additional in-plane gates and a top gate (not labeled in Fig.~\ref{fig2}(a)) control the coupling of the ring to source and drain contacts. Experiments were carried out in a dilution refrigerator with a base temperature of 80~mK using dc and low-frequency ac lock-in techniques.

\section{Zeeman splitting and ground state spins}
\begin{figure}[tbph]
\centering
\includegraphics[width=14cm]{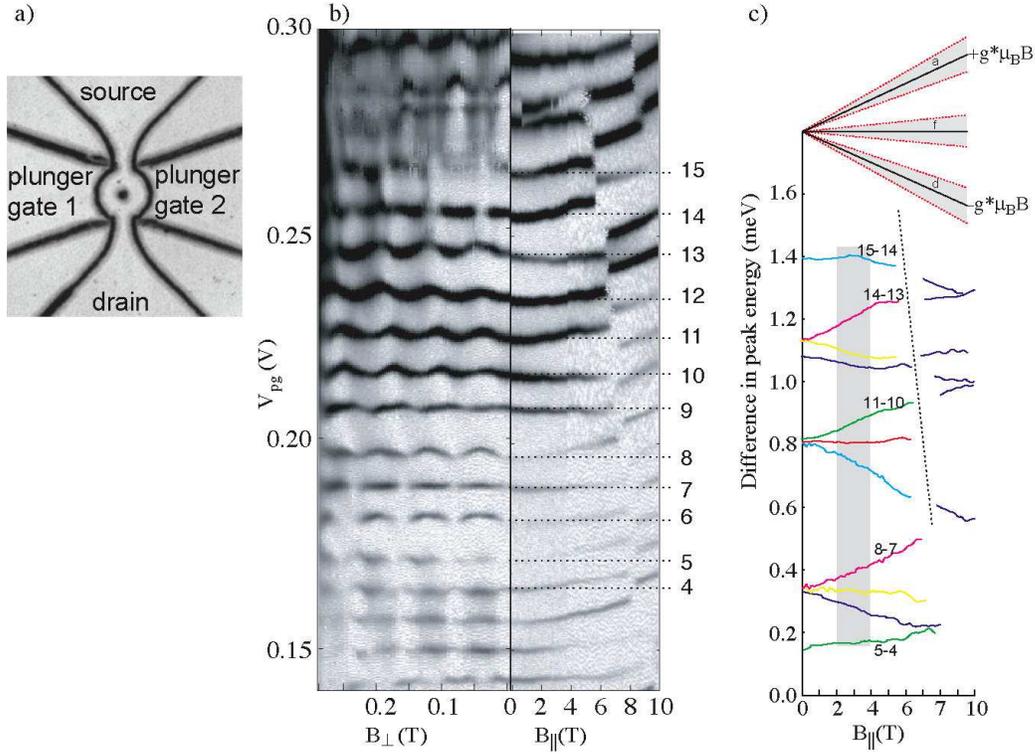}
  \caption{(a) SFM image of the sample. Dark lines represent the oxide lines. The quantum ring is coupled to source and drain via quantum point contacts in the tunneling regime. Plunger gates left and right of the ring tune the electron number in the Coulomb-blockade regime of the ring one by one. (b) From left to right: Spectrum as a function of perpendicular magnetic field and as a function of in-plane magnetic field. The peaks for the in-plane field measurement were all offset by the same small amount in order to match the spectrum before rotation of the sample. (c) Peak spacing of neighboring conductance peaks from the parallel field data in (b). The three expected slopes labeled a (ascending), f (flat), and d (descending) are shown at the top.}
\label{fig2}
\end{figure}
For the observation of clear Zeeman splitting in a quantum dot it is crucial to eliminate orbital effects of the magnetic field that can be orders of magnitude larger. The magnetic field can be applied normal or parallel to the plane of the electron gas.
Figure~\ref{fig2}b shows the conductance of the ring (gray scale) as a function of plunger gate voltage $V_\mathrm{pg}=V_\mathrm{pg1}=V_\mathrm{pg2}$ (vertical axis) and magnetic field (horizontal axis). The latter was applied normal to the plane of the electron gas for the left panel and parallel for the right panel. In both cases, conductance peaks are visible as dark stripes, separated by bright regions of very small conductance. Conductance peaks are numbered consistently throughout this paper. If the magnetic field is oriented normal to the plane of the ring, conductance peaks show Aharonov--Bohm-type oscillations periodic with each additional flux quantum $h/e$ penetrating the ring \cite{fuhrer01,ihn04}. States that move up and down in plunger gate as a function of magnetic field are extended around the ring, while those with a weak magnetic field dispersion tend to be localized in one ring arm \cite{fuhrer01,ihn04}. Neighboring peaks oscillating exactly in parallel are strong candidates for spin pairs \cite{luscher01,ihn03}, i.e. the same orbital state is successively occupied by spin-up and spin-down. An example of such a pair is peak 10 and 11. Raw data showing this pair together with its Zeeman splitting was published in Ref.~\cite{ihn03}.

When the magnetic field is oriented parallel to the plane of the ring (right panel in Fig.~\ref{fig2}(b)), individual conductance peaks show a weak parabolic diamagnetic shift to higher plunger gate voltages. Considering the different magnetic field scales for the two orientations it is apparent from Fig.~\ref{fig2}(b) that the diamagnetic shift of levels in $B_\parallel$ is about an order of magnitude smaller than the level shift in $B_\perp$. In addition it is found here, in agreement with previous investigations \cite{lindemann02,weis94}, that the diamagnetic shift of levels is, with sufficient accuracy,  the same for neighbouring conductance peaks. This property allows the elimination of the diamagnetic shift, if peak spacings are analyzed rather than absolute peak positions. This is necessary because the expected Zeeman splitting in GaAs is tiny compared to the orbital effects discussed so far: an estimate using the GaAs bulk g-factor gives $g\mu_\mathrm{B}=25$~$\mu$eV/T, where $\mu_\mathrm{B}$ is Bohr's magneton. The peak spacings determined from the parallel field data in Fig.~\ref{fig2}(b) are shown in Fig.~\ref{fig2}(c). These peak spacings can be interpreted using \cite{lindemann02}
 \[ e\alpha_\mathrm{pg}\Delta V_\mathrm{pg}^{(N+1)}(B_\parallel) = (s_{N+1}-2s_N+s_{N-1})g\mu_\mathrm{B}B_\parallel+\mathrm{const.}, \]
 where $s_N$ is the spin quantum number of the dot along $B_\parallel$ and $\alpha_\mathrm{pg}$ is the lever arm of the plunger gate. Differences of successive ground state spins $\delta_N=s_N-s_{N-1}$ take on half-integer values that can be interpreted as the `spin of the added electron'. The second difference $\zeta_{N}=s_{N+1}-2s_N+s_{N-1}$ takes on integer values. The slopes of peak spacings observed in Fig.~\ref{fig2}(c) can be classified in the range between 2 and 4~T (shaded region) by the three integer values -1, 0 and 1 as descending, flat and ascending. The results of an experimental evaluation of the slopes $g\zeta_N$ and the $\zeta_N$ values are tabulated in Table~\ref{tab1}. Some of the slopes, e.g., (11-10) and (6-5), are in excellent agreement with the GaAs bulk g-Factor of -0.44, and most of them fall into the gray shaded areas d,f,a indicated at the top of Fig.~\ref{fig2}(c). The only exceptions are (5-4) and (13-12). The origin of these fluctuations in the slopes that are also reported by other authors \cite{folk01} remain to be investigated. In our case, part of the fluctuations could arise due to small differences in lever arms for different states ($<10\%$) or due to other states with different spin that are close in energy.
 
From the knowledge of a series of values $\zeta_N$, series of ground state spins can be generated. Table~\ref{tab1} shows the ground state spins $s_N$ and the added electron spins $\delta_N$ for the data shown in Fig.~\ref{fig2}(c). It is reassuring to notice that the $\delta_N$ values are in agreement with two families of $B_\parallel$-dispersions in the raw data in Fig.~\ref{fig2}(b). Peaks with $\delta_N=1/2$ tend to curve upwards, while those with $\delta_N=-1/2$ tend to stay flat. This direct analysis is possible, because the diamagnetic shift of peaks is small in our sample. The $s_N$-series in Table~\ref{tab1} is not unique. Other series can be generated by adding a constant integer or half integer to all the given $s_N$ values. Additional information is therefore required to pinpoint the ground state spins uniquely. It is known from numerical calculations of quantum dots in a similar regime \cite{jiang03} that ground state spins with $|s_N|>3/2$ are extremely unlikely. It turns out that this restriction reduces the number of possible ground state spin series to only four, i.e. the one given in Table~\ref{tab1} and those where 1/2, 1 or 3/2 is added to all the given $s_N$-values.
\begin{table}[tbhp]
 \begin{tabular}{l|ccccccccccccc|}
 peak          &   & 4      &        & 5    &     & 6    &       & 7    &     & 8      &        & 9    &    \\ 
 \hline
 $g\zeta_N$& &         & 0.18&      &-0.44&   & -0.13&      &0.5&       &-0.55&        &0.01\\
$\zeta_N$ &    &         & 0    &        &-1 &        & 0   &        &  1&         & -1     &       & 0 \\
$\delta_N$&    & 1/2 &        & 1/2 &    &-1/2 &       &-1/2 &     &  1/2 &        &-1/2 &    \\
$s_N$        & -1 &        & -1/2 &        & 0 &       & -1/2 &       & -1  &         & -1/2 &        & -1 \\
\hline
 \end{tabular}
 
 \begin{tabular}{l|ccccccccccc|}
peak           & 10  &        & 11  &    & 12    &        & 13 &      & 14    &     & 15      \\
\hline
$g\zeta_N$&       &0.42&       &-0.17&     &-0.26&      &0.52&       &-0.12&        \\
$\zeta_N$ &        &  1    &       & 0 &         & -1    &       &  1 &         & 0   &                 \\
$\delta_N$& -1/2 &        &  1/2&    &  1/2 &        & -1/2 &     & 1/2 &       &  1/2      \\
$s_N$        &        & -3/2 &        & -1 &         & -1/2 &        & -1 &         & -1/2 &          \\
\hline
\end{tabular}
\caption{Ground state spins $s_N$ and added electron spins $\delta_N$ for the data shown in Fig.~\ref{fig2}(c). The values $g\zeta_N$
were determined from average slopes of lines in Fig.~\ref{fig2}(c) between 2 and 4~T.}
\label{tab1} 
\end{table}

\section{Spin-paired orbital levels}
\begin{figure}[tbp]
\centering
\includegraphics[width=12cm]{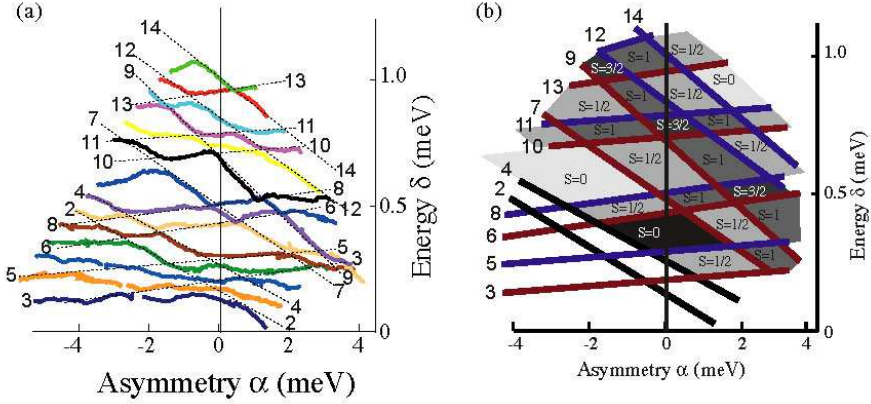}
  \caption{\small (a) Energy levels in the ring as a function of asymmetry. Plotted are the positions of conductance peaks with an energy of 270~$\mu$eV subtracted between neighboring peaks. (b) Corresponding spin configuration as described in the text.}
\label{fig3}
\end{figure}

Additional information about the states in the ring can be obtained by applying a finite voltage difference between the two plunger gates to
the left and the right of the ring \cite{fuhrer03,ihn03a}. The level spectrum plotted as a function of this `asymmetry' $\alpha$ is shown in Fig.~\ref{fig3}(a).
Two types of states can be recognized: states that are extended around the ring have a weak dependence on $\alpha$ (e.g., states 6,8,10,11) while those localized in one arm of the ring depend strongly on $\alpha$ (e.g., states 4,7,9). This spectrum allows, together with the $B_\perp$-data in Fig.~\ref{fig2}(b), to identify spin pairs. For example, the extended states 10 and 11 run exactly in parallel as a function of $\alpha$ (Fig.~\ref{fig3}(a)) and as a function of $B_\perp$ (Fig.~\ref{fig2}(b)) and are therefore identified as a spin pair. The same is true for extended states 6 and 8 which do not lead to neighboring conductance peaks at zero asymmetry. It can be verified in the $\delta_N$-row of Table~\ref{tab1} that each of these spin pairs indeed has one spin-up (1/2) and one spin-down (-1/2) assigned for the tunneling electron. Among the more localized states, 2 and 4 can be called a spin pair owing to their identical slope in Fig.~\ref{fig3}(a) and their $B_\perp$ behavior in Fig.~\ref{fig2}(b). In contrast, states 7 and 9 have different slopes (Fig.~\ref{fig3}(a)), the same spin (Table~\ref{tab1}) and are therefore not a spin pair. However, 7 pairs up with 12 and probably also 9 with 14.
All the spin pairs have the property that $\delta_N=-1/2$ is occupied before $\delta_N=1/2$, because the former is lower in energy. The only remaining unpaired level in the range of our measurement is 5. Taking the typical spacing between spin pairs into account and considering that state 5 has $\delta_N=1/2$, we conclude that the missing
partner for state 5 is lower in energy. It is therefore reasonable to assume that all states below 5 are paired and that therefore the ground state spin in valley (6-5) is zero. With this assumption, the ground state spin series given in Table~\ref{tab1} is the one that we favorize among the four possible. Figure~\ref{fig3}b shows the resulting ground state spin states for the data in Fig.~\ref{fig3}(a).

\section{Exchange interactions and singlet--triplet transition}
\begin{figure}[tbp]
\centering
\includegraphics[width=12cm]{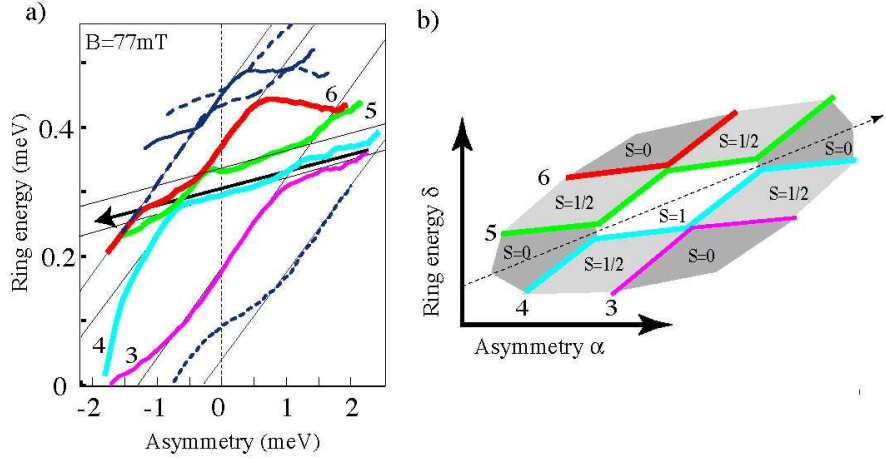}
  \caption{\small (a) Energy levels in the ring as a function of asymmetry. Plotted are the positions of conductance peaks with an energy of 270~$\mu$eV subtracted between neighboring peaks. Data are adapted from Ref.~\cite{fuhrer03}. (b) Corresponding spin configuration.}
\label{fig4}
\end{figure}

It is interesting to note that the phase diagram in Fig.~\ref{fig3}(b) predicts transitions from $s_N=0$ to $s_N=1$ at constant electron number,
driven by gate voltages. Such singlet--triplet transitions are interesting in view of the Hamiltonian (\ref{ham}) because the triplet state becomes
the ground state due to exchange interactions. The quantum ring with its states classified as `extended' or `localized' is in certain respects comparable
to a double quantum dot system with states, e.g., classified as `left' or `right'. It has been shown that the magnitude of exchange matrix elements
can be extracted from such experiments \cite{tarucha00,fuhrer03}. It turns out, however, that in the data shown above, the exchange energy is smaller
than the broadening of energy levels and cannot be observed. However, in another cooldown of the same sample, a gate-voltage induced singlet--triplet transition was identified and the exchange energy was extracted \cite{fuhrer03}. Figure~\ref{fig4}a shows the measured energy spectrum as a function of plunger gate asymmetry $\alpha$ with localized states (steep ascending slope) and extended states (weakly ascending slope). The spectrum can be interpreted as the crossing of two spin-paired levels. The corresponding spin filling extracted from a similar analysis as that discussed above is depicted in Fig.~\ref{fig4}(b). Tuning the system with gate voltages along the line indicated by the dashed arrow leads to a singlet--triplet--singlet transition for the
uppermost two electrons. The extracted exchange energy matrix element of 25~$\mu$eV is larger than $kT$, but about an order of magnitude smaller than the corresponding Hartree energies for extended states (360~$\mu$eV), localized states (530~$\mu$eV) and the Hartree energy matrix element for a pair of an extended and a localized state (335~$\mu$eV) that can be extracted \cite{fuhrer03}. Singlet--triplet transitions have been observed in strong magnetic fields before \cite{tarucha00,ciorga01,kyriakidis02}.

In conclusion, we have shown how the spin states of a Coulomb blockaded many-electron quantum ring system can be investigated in great detail.
Zeeman splitting of energy levels was observed in an in-plane magnetic field and possible series of spin ground-states were found. Using the energy shift of orbital states as a function of perpendicular magnetic field and asymmetric gate voltages, pairs of states with identical orbital wave functions but opposite spins were identified. Using this knowledge, the number of plausible sequences of ground state spins could be reduced to only one. Knowing the spin ground states of the ring as a function of experimental parameters, it was possible to identify a gate-voltage induced singlet--triplet transition. The magnitude of the exchange matrix element could be extracted.

\end{document}